\documentclass[%
 aip,
 amsmath,amssymb,
 reprint,%
]{revtex4-1}
\usepackage{CJK}
\usepackage{graphicx}
\usepackage{bm}

\usepackage{afterpage}
\usepackage[utf8]{inputenc}
\usepackage[T1]{fontenc}
\usepackage{newtx}
\usepackage{etoolbox}
\DeclareMathAlphabet{\mathcal}{OMS}{cmsy}{m}{n}

\makeatletter
\def\@email#1#2{%
 \endgroup
 \patchcmd{\titleblock@produce}
  {\frontmatter@RRAPformat}
  {\frontmatter@RRAPformat{\produce@RRAP{*#1\href{mailto:#2}{#2}}}\frontmatter@RRAPformat}
  {}{}
}%
\makeatother

\begin{document}

\preprint{AIP/123-QED}

\title[]{Distinct Lifetime Scaling Laws of Turbulent Puff in Duct Flow}
\author{Jiashun Guan (\begin{CJK*}{UTF8}{gbsn}关家顺\end{CJK*})}
\author{Jianjun Tao (\begin{CJK*}{UTF8}{gbsn}陶建军\end{CJK*})$^*$}%
 \email[Author to whom correspondence should be addressed:]{ jjtao@pku.edu.cn}
\affiliation{ HEDPS-CAPT, SKLTCS, Department of Mechanics and Engineering Science, College of Engineering, Peking University, Beijing 100871, China
}%


\begin{abstract}
The spatio-temporal dynamics of localized turbulent puffs --- the characteristic transitional structures in square duct flows --- are investigated through direct numerical simulations and theoretical analyses. It is revealed that the turbulent puffs are transient structures, exhibiting distinct relaminarization regimes  bifurcated at a critical Reynolds number $Re_c\simeq1450$. Puff's mean lifetimes at the subcritical regime ($Re<Re_c$) follow a square-root scaling law with increasing $Re$, transitioning to a super-exponential scaling in the supercritical regime ($Re > Re_c$). By implementing pattern preservation approximation, the Reynolds-Orr kinetic energy equation is reduced to a noisy saddle-node bifurcation equation, which explains the observed scaling laws in terms of the deterministic decay governed by the critical slowing down at the subcritical regime, and the abrupt decay activated by the stochastic fluctuations. Despite  geometric confinement inducing unique secondary flows, e.g., corner-localized streamwise vortex pairs, corner-aligned high-speed streaks, and forked low-speed streaks,  the puff lifetime statistics remain analogous to those in pipe flows, suggesting geometric invariance in decay mechanisms for transitional wall-surrounded turbulence.

\end{abstract}

\maketitle

Similar to pipe flows \cite{Reynolds1883, Eckhardt2007,Mullin2011}, the transition to turbulence in square ducts remains an unresolved challenge in fluid dynamics. Square duct flows exhibit linear stability \cite{Tatsumi1990,Demyanko13}, requiring finite-amplitude perturbations to trigger the subcritical laminar-turbulent transition. Key features of duct turbulence include secondary mean motions, such as low-Reynolds-number vortex secondary structures \cite{Nikuradse1926, uhlmann2007}, deformations of mean velocity profile \cite{Gavrilakis1992}, and flow modifications caused by corner effect \cite{Huser1993}. While numerical studies have quantified Reynolds number effects on secondary flows \cite{Pinelli2010}, existing analyses focus predominantly on short periodic ducts. Recent researches of elongated ducts reveal localized turbulent puffs as the characteristic transitional structures \cite{Takeishi2015,Barkley2015,guan2024}, with puff and slug formations investigated by introducing inlet perturbations \cite{Khan2021}. However, the mean-flow vortex characteristics of the turbulent puffs in duct geometries remain unexplored.

A central question about the subcritical transitions is whether the characteristic transitional structures exhibit self-sustaining behavior or possess finite lifetimes. Extensive studies of pipe flow have demonstrated that  turbulent puffs \cite{Wygnanski1973} represent transient phenomena \cite{Hof2006}: their lifetime statistics follow exponential distributions \cite{Peixinho2006, Eckhardt2007, Avila2010}, while their  mean lifetimes display  super-exponential growth as the Reynolds number increases  \cite{Hof2008, Avila2011}. At elevated Reynolds numbers, puff splitting emerges, with  turbulence sustenance governed by the balance between splitting and decay processes \cite{Wygnanski1975,Avila2011}. Phenomenological models \cite{Barkley2016,Shih2016} and first principles model \cite{guan2025} have been developed to capture macroscopic puff dynamics. In contrast to the well-characterized pipe flow system, the lifetime statistics of turbulent puffs in duct flows have not been studied so far, and it is still unknown whether the turbulent puffs can be self-sustained in the presence of secondary flows. 

\begin{figure}[h]
	\centering
	\includegraphics[width=\linewidth]{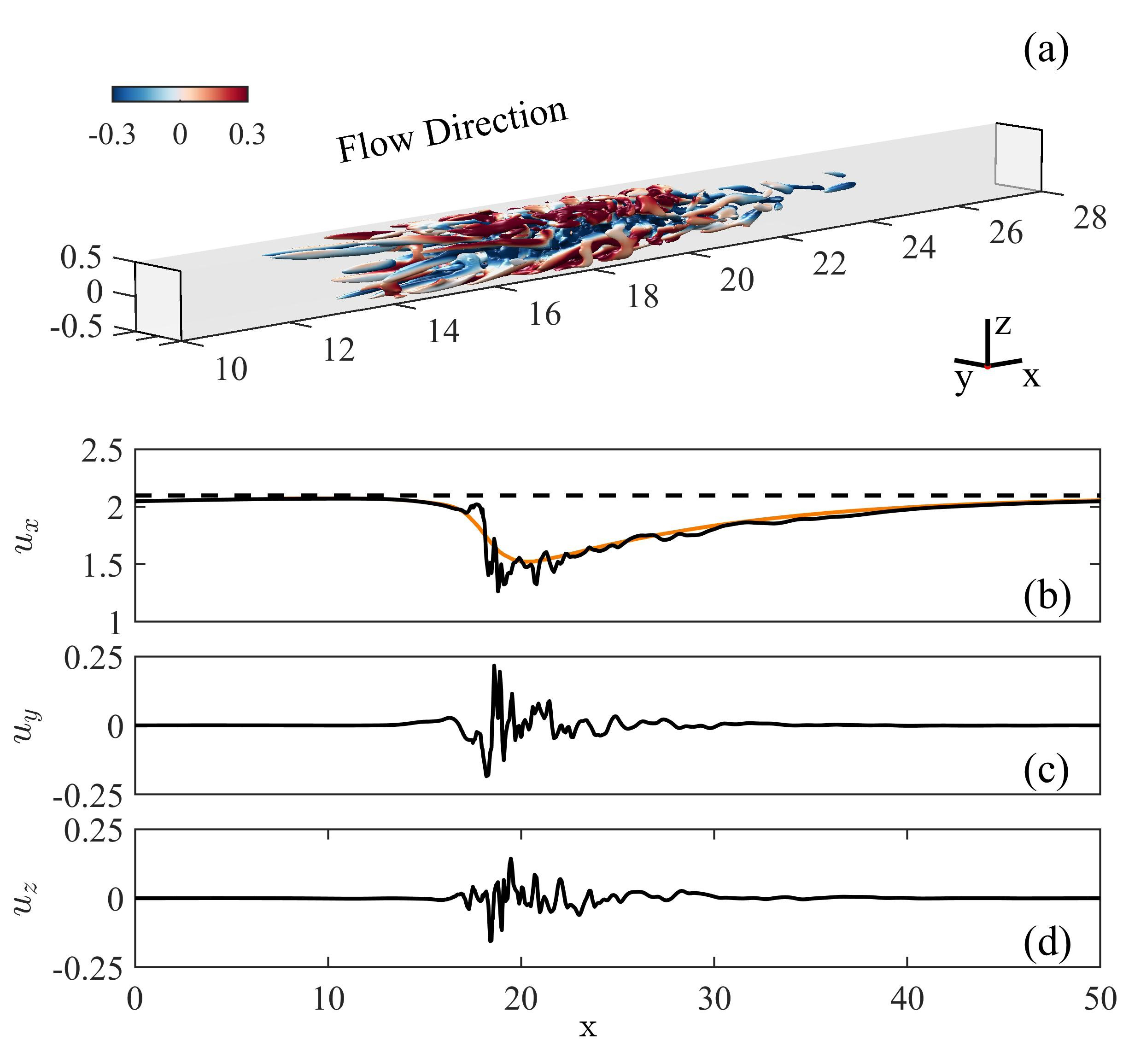}
	\caption{Flow field of a turbulent puff obtained at $Re=1510$. (a)  Iso-surfaces of the vortex criterion $Q=0.0125$, calculated with the disturbance field and colored by the streamwise disturbance velocity $u^{\prime}_x$. Velocities $u_x$, $u_y$, and $u_z$ at the duct centerline  are shown in (b), (c) and (d), respectively. In (b), the dashed line and the red line indicate the basic flow velocity $U_x =2.097$ and the ensemble average of  $u_x^{\prime}$ at the centerline, respectively. }
	\label{fig:f1}
\end{figure}

We consider the flow in a long and straight duct with the width $H$ of the square cross-section and the bulk velocity $U$ chosen as the characteristic length and velocity, respectively. The origin of the Cartesian coordinates $(x, y, z)$ is located at the center of the duct with $x$ defined along the streamwise direction. The incompressible Naiver-Stokes (NS) equations with velocity components $u_x, u_y$, and $u_z$ (along directions $x, y$, and $z$, respectively) are solved with the open-source spectral element code Nek5000 \cite{nek5000_webpage_2008},  and no-slip boundary conditions on the walls, streamwise periodic condition for flow field, and constant flow rate are implemented. Spectral elements are uniformly distributed in the streamwise direction and bunched towards side walls to adapt large velocity gradients near the walls. Locations of element vertices in the $y$ and $z$ directions are distributed in the same mapping:
\begin{equation}
    y_n = \ z_n=\frac{tanh[\varepsilon(2h_n-1)]}{2tanh(\varepsilon)}
\end{equation}
where $h_n=n/N$ with $n=0, 1, 2, ... N$,  and the stretching parameter $\varepsilon$ is set as 1.2.  Direct numerical simulations (DNS) are performed using a mesh of $10\times10$ elements in the $y$-$z$ plane and 100 elements over a duct length of $L=50$, resulting in $10^4$ elements in total. The time step $\delta t$ for the third-order Backward Difference Formula \cite{nek5000_webpage_2008} is $0.005$, corresponding to a CFL number of approximately 0.5. Within each element, a seventh spectral order ($n^{th} =7$) is applied. At $Re=1540$, we verify that the mean volume-integral disturbance kinetic energies ($\bar{E_k}$) for configurations $(n^{th}, \delta t, L)=(7, 0.005, 50)$ and (9, 0.003, 50) differ by less than $3\%$. Similarly, configurations $(n^{th}, \delta t, L)=(7, 0.005, 50)$ and (7, 0.005, 100) with double streamwise elements (200) yield a relative difference in $\bar{E_k}$ of less than $3\%$. Therefore, the present DNS configurations are validated as sufficiently accurate to capture the main characteristics of puffs in the explored parameter space. The Reynolds number is defined as $Re=UH/\nu$ with $\nu$ as the kinematic viscosity of the fluid. 

The basic flow is assumed to be steady and parallel, leading to a reduction of the NS equations to $\Delta_x U_x=Re\nabla_x p$, where $\Delta_x$ denotes the Laplacian in the cross-section, $\nabla_x p=dp/dx$, and $Re\nabla_x p$ is a constant ensuring the constant flow rate condition, i.e., the integral within the cross-section $\int U_x ds=1$. Note that $U_0=(Re\nabla_x p/2)(z^2-1/4)$ satisfies the NS equation and the boundary conditions at $z=\pm1/2$, and then the complete solution is assumed in the form of $U_x=U_0+U_1$, where $U_1$ satisfies $\Delta_xU_1 = 0$. Applying the no-slip boundary conditions and separation of variables, $U_1$ can be solved, yielding the basic flow solution:
\begin{subequations}
\begin{equation}
            U_x(y,z) = \frac{Re\nabla_x p}{2} (z^2-\frac{1}{4}) + \frac{4 Re\nabla_x p }{\pi^3} \hat{u},
\end{equation}
\begin{equation}
    \hat{u}=\sum_{k=1}^{\infty} \frac{\cosh\left[ (2k-1)\pi y \right]}{(2k-1)^3 \cosh\left[ \frac{(2k-1)\pi }{2} \right]} \sin\left[ \pi(2k-1)(z+\frac{1}{2}) \right],
\end{equation}
\begin{equation}
    U_y(y,z)=U_z(y,z)=0.
\end{equation}
\end{subequations}
 In Eq. (2), $y$ and $z$ can be exchanged due to the symmetric property of the basic flow. The disturbance velocity $ \mathbf{u^{\prime}} = \mathbf{u}-\mathbf{U} $, and the volume-integral disturbance kinetic energy $E_k=\frac{1}{2}\int_{V}u_{i}^{\prime}u_{i}^{\prime} dV$ satisfies the Reynolds-Orr equation due to the boundary conditions,
\begin{equation}
	\frac{dE_k}{dt}=\mathcal{P}-\frac{ \mathcal{D}}{Re}=-\int_{V} u_{i}^{\prime} u_{j}^{\prime} \frac{\partial U_{i}}{\partial x_{j}} dV-\frac{1}{Re}\int_{V} \frac{\partial u_{i}^{\prime}}{\partial x_{j}} \frac{\partial u_{i}^{\prime}}{\partial x_{j}} dV
\label{eq:pdek}
\end{equation}
where $ \mathcal{P}$ and $ \mathcal{D}/Re$ represent the production and dissipation terms, respectively. 

At a moderate Reynolds number, finite-amplitude localized perturbations can trigger turbulent patches, which evolve into developed turbulent puffs [Fig.~\ref{fig:f1}(a)] with statistically constant streamwise length and convection velocity, surrounded by laminar regions \cite{Barkley2015}. The vortex structures of puff emerge from the near wall regions at the upstream front, and gradually decay at the center region of puff's downstream tail, with strong velocity fluctuations lying near the upstream front as shown in Fig. ~\ref{fig:f1}(b)-(d). These features are similar to those of puffs in  pipe flows \cite{willis2007,song2017}. 

\begin{figure*}
	\includegraphics[width=0.83\linewidth]{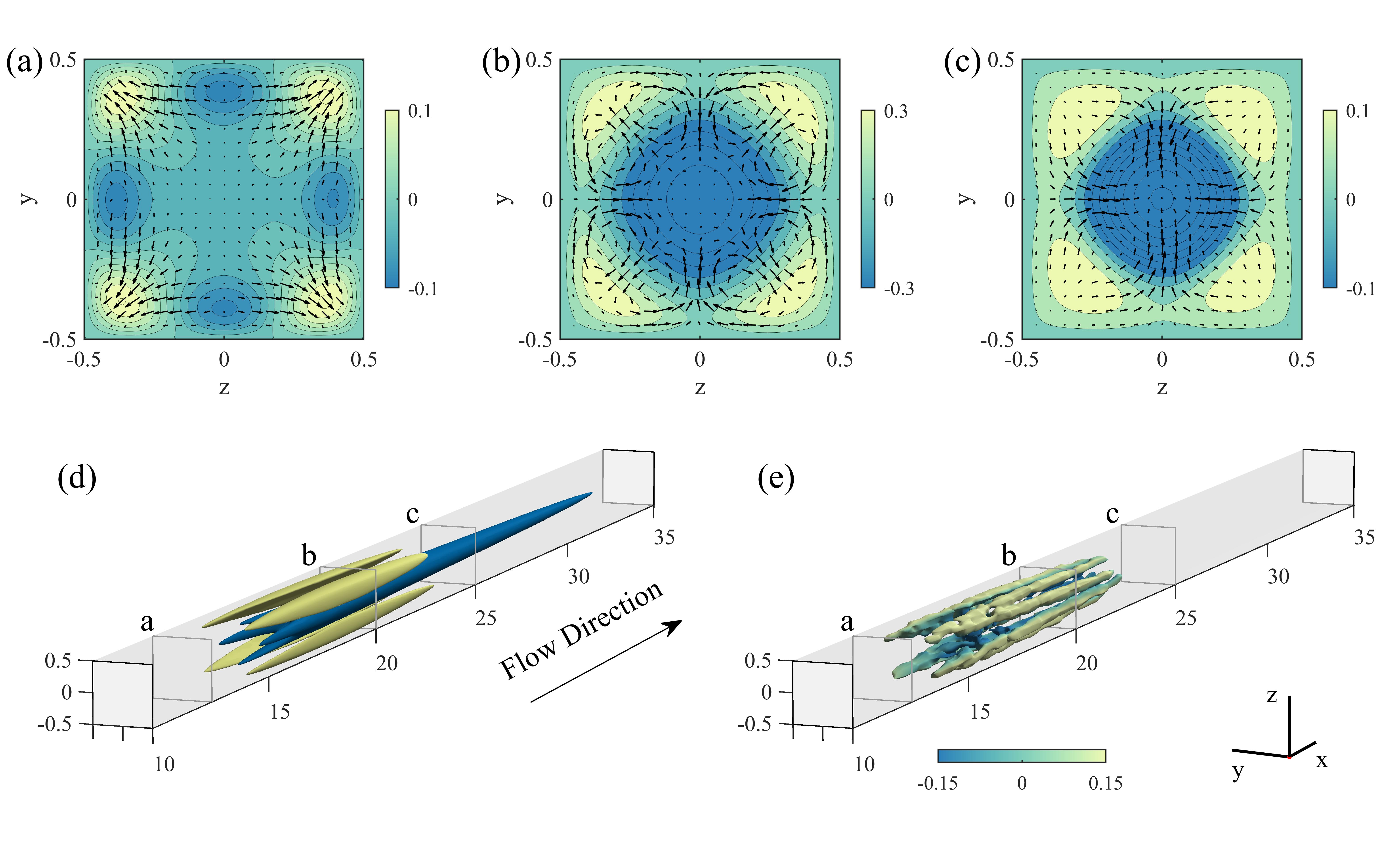}
	\caption{The ensemble-averaged disturbance flow field $\langle\mathbf{u'}\rangle$ of puffs obtained at $Re=1510$. The velocity fields in the cross-sections at $x=12.5,\ 20\ (\text{the centroid}),$ and 25 are shown in (a), (b), and (c), respectively, with the iso-contours of $\langle u_x^{\prime}\rangle$. (d) Iso-surfaces of  $\langle u_x^{\prime}\rangle=-0.2$ (blue) and 0.2 (yellow). (e) Iso-surfaces of the vortex criterion $Q=0.001$, calculated with $\langle\mathbf{u'}\rangle$ and colored with $\langle u_x^{\prime}\rangle$.}
	\label{fig:f2}
\end{figure*}

By defining the puff centroid $x_c$ with the disturbance kinetic energy $E_k$ as $x_c={\int_V E_k xdx}/{\int_V E_k dx}$,  the ensemble averaged flow of puff at $Re=1510$ is calculated by shifting the centroids of $1945$ DNS fields chosen from the time series (totally lasting $10^4$) to $x=20$. As shown in Fig. ~\ref{fig:f2}, the large scale flow structures of the averaged field exhibit distinct characteristics at different streamwise locations. At the middle part,  four streamwise vortex pairs exhibit at the corners (Fig. 2b, 2e), similar to the secondary flow of turbulence in short periodic ducts \cite{Gavrilakis1992,Pinelli2010}. Each pair of the streamwise vortex structures combines at the upstream region (Fig. 2e), corresponding to a jet flow  in the cross section rushing towards the corner (Fig. 2a),  and tilts to the center at the downstream side. At the upstream front, the main features are the high speed streaks at the corners (Fig. 2a, 2d) and low speed streaks near the side wall centers, which merge with each other at the middle part to form a central low-speed streak, elongating into the downstream side as shown in Fig. 2(d).  The locations of the high and low speed streaks are deterministic in the ensemble averaged flow field due to the duct geometric confinement, and the streak structures look similar to those of relative periodic orbits (temporally periodic flow in a frame comoving at a constant speed) found in pipe flows \cite{Avila2013}.   

\begin{figure*}
    \includegraphics[width=0.8\linewidth]{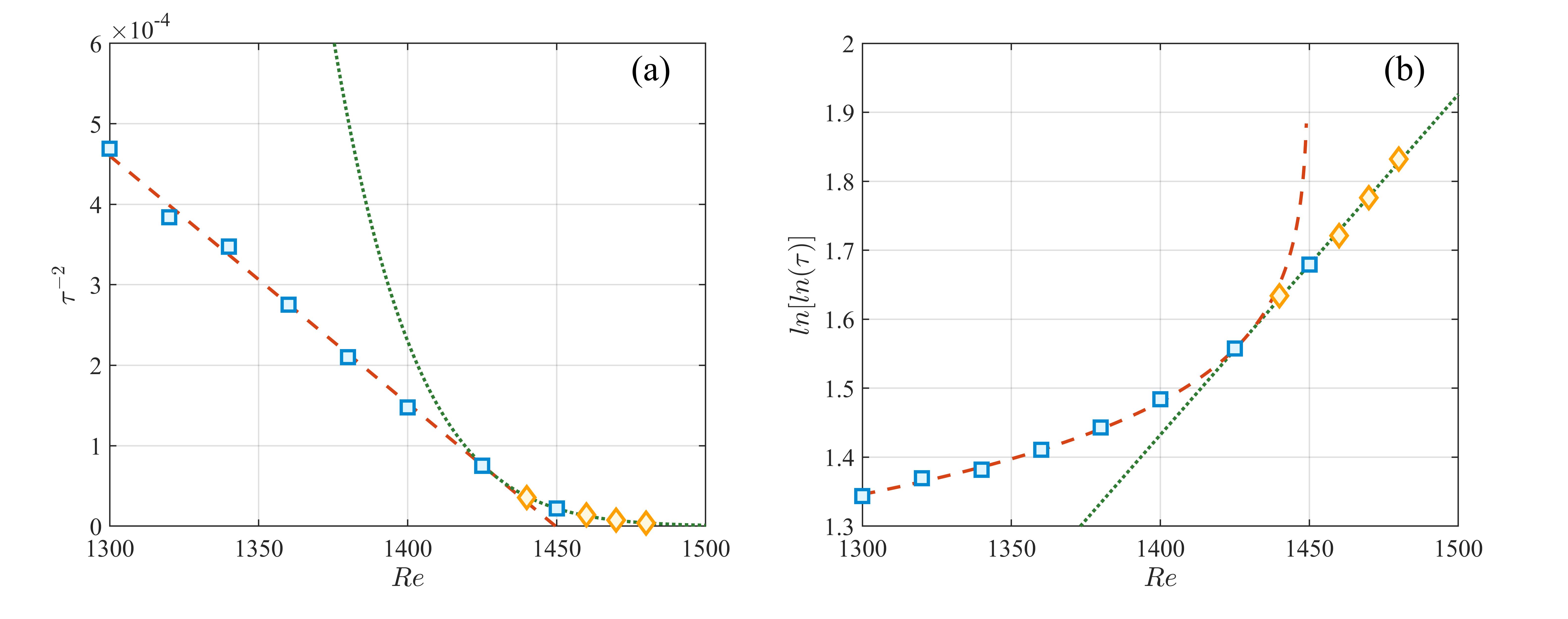}
    \caption{Mean lifetime $\tau$ as a function of $Re$ with (a) $\tau^{-2}$ and (b) $ln[ln(\tau)]$ as the ordinate. The blue squares and yellow diamonds represent the results with  $Re_{ini}=1500$ and 1510, respectively. The dashed and solid curves stand for Eq. (4) and (5), respectively.  }
    \label{fig:f3}
\end{figure*}

Next, extensive simulations are implemented to analyze the statistical properties of puff lifetime. As shown in Fig. 3, each square symbol represents the mean lifetime of 100 DNS cases initialized using 100 instantaneous fields (separated by at least 5 time units) from a developed puff obtained at $Re_{ini} =1500$, while each diamond symbol denotes the mean value of 50 DNS cases initialized with puffs at $Re_{ini} =1510$. The total number of DNS cases in Fig. 3 is 1000.  When $E_k<0.05$, it is found that small-scale fluctuations disappear and $E_k$ decays monotonically. Consequently, the relaminarization criterion is set as $E_k = 0.05$, and it is checked that the main findings in this letter are not sensitive to this choice. The mean lifetime is defined as $ \tau(Re)=(\sum_{i=1}^{M}\tau_i)/M$, where $M$ represents the total number of DNS cases at $Re$, and the lifetime of each case $\tau_i$ is counted from the initialization to the time meeting the relaminarization criterion. It is shown in Fig.~\ref{fig:f3}(a) that the mean lifetime satisfies a -2 scaling law at low Reynolds numbers
 \begin{equation}
 	\tau=(a_1Re+b_1)^{-1/2},
 \end{equation}
 where the coefficients $(a_1, b_1)=(-3.08\times10^{-6}, 4.46\times10^{-3})$ are fitted with data as $Re<1440$. This scaling  suggests a critical Reynolds number for infinite lifetime of puff, i.e. $Re_c=-b_1/a_1=1450$. Such a square-root scaling law at low Reynolds numbers has been found for the mean lifetime of localized wave packet in two-dimensional channel flows \cite{Zhang2022} and puffs in pipe flows \cite{guan2025}.

When $Re>1450$, however, the mean lifetimes of puffs still remain finite values as shown in Fig.~\ref{fig:f3}(b), and satisfies a super-exponential scaling as:
\begin{equation}
    ln[ln(\tau)]=a_2Re+b_2,
\end{equation}
where the coefficients $(a_2, b_2)=(4.94\times10^{-3}, -5.48)$ are fitted with data as $Re>1450$, indicating that the turbulent puffs in duct flows are transient structures. The square-root scaling and the super-exponential scaling of puff's mean lifetime, to the best of our knowledge, have not been reported for duct flows. For pipe flows, it is found that the puffs are transient  \cite{Hof2006}  and their mean lifetimes follow the super-exponential scaling  \cite{Hof2008, Avila2011}. Recently, the transition from square-root scaling to super-exponential scaling of mean lifetime is revealed for puffs in pipe flows, representing a switch from the deterministic decay of kinetic energy to the memoryless decay, governed by a noisy saddle-node bifurcation \cite{guan2025}. Consequently, it is necessary to study the underlying mechanisms governing the scaling laws and the physical meaning of the critical Reynolds number $Re_c$ for duct flows.

\begin{figure*}[t]
    \centering
    \includegraphics[width=0.97\linewidth]{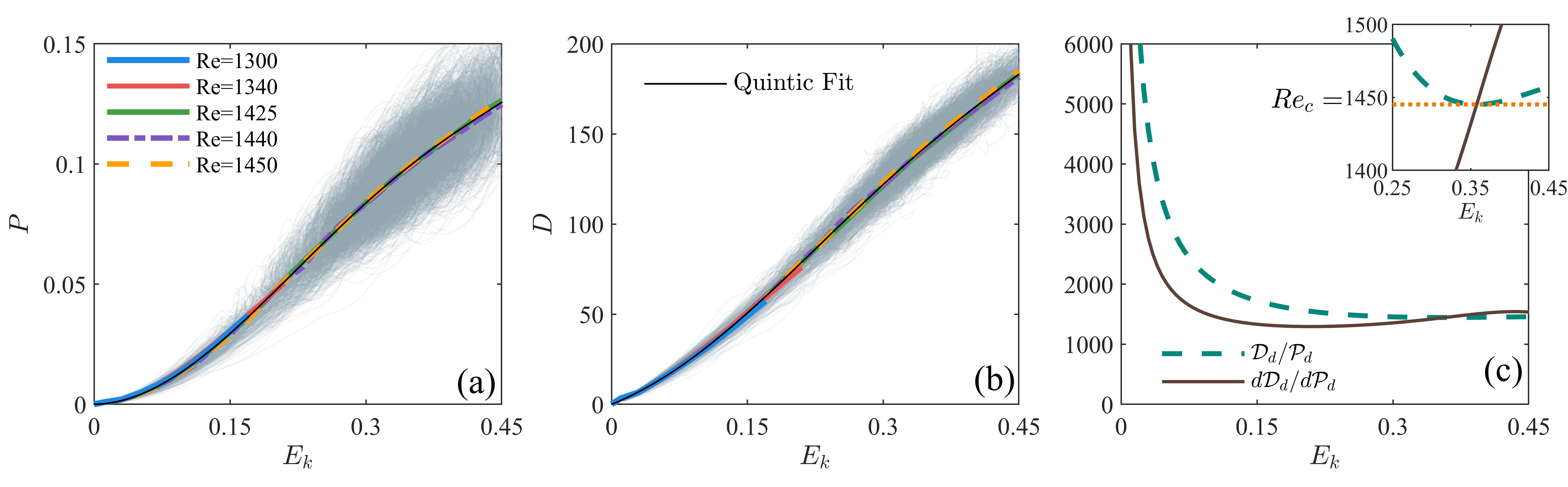}
    \caption{Trajectories of puffs in the (a) $ \mathcal{P}-E_k$ and (b) $\mathcal{D}-E_k$ phase spaces. The ensemble averages at each $E_k$ are shown as thick color curves and are fitted with quintic polynomials (black solid curves). (c) $ \mathcal{D}_d/ \mathcal{P}_d$ and $d \mathcal{D}_d/d \mathcal{P}_d$ as functions of $E_k$. }
    \label{fig:f4}
\end{figure*}
 
The production and dissipation terms in Eq. (3) may be decomposed into  stochastic components ($\mathcal{P}_s$ and $\mathcal{D}_s$) and  deterministic components ($\mathcal{P}_d$ and $\mathcal{D}_d$) as   
 \begin{equation}
 	\frac{dE_k}{dt}= \mathcal{P}_d-\frac{ \mathcal{D}_d}{Re}+ (\mathcal{P}_s-\frac{ \mathcal{D}_s}{Re})=\sigma_d+\sigma_s,
 \end{equation}
 where $\sigma_d= \mathcal{P}_d-{\mathcal{D}_d}/{Re}$ and $\sigma_s$ represent the growth rates of disturbance kinetic energy  contributed by the deterministic and stochastic components within a puff, respectively. The temporal evolution of puff corresponds to the trajectories in $ \mathcal{P}-E_k$ and $\mathcal{D}-E_k$ phase spaces, and   $\mathcal{P}_d$ and $\mathcal{D}_d$ are defined as the ensemble averages of $ \mathcal{P}$ and $\mathcal{D}$ at given $E_k$ slices with a width of 0.02, illustrated by curves  in Fig.~\ref{fig:f4}(a) and (b), respectively. It is checked that the variations of $\mathcal{P}_d$ and $\mathcal{D}_d$ caused by changing the slice width to 0.01 can be ignored. For the subcritical transition in two-dimensional channel flows, where the stochastic components $\mathcal{P}_s$ and $\mathcal{D}_s$ can be ignored, a pattern preservation approximation is validated at low and moderate Reynolds numbers: $\mathcal{P}_d$ and $\mathcal{D}_d$ in the Reynolds-Orr equation are only functions of $E_k$, and are nearly independent of $Re$ \cite{Zhang2022}.  This approximation can be extended for the duct flows: it is shown in Fig.~\ref{fig:f4} that the $\mathcal{P}_d$ and $\mathcal{D}_d$ curves obtained at different $Re$ almost coincide with each,  depicted well by the quintic fitting curves. For cases with low Reynolds numbers, puffs' kinetic energies seldom reach high values, and hence only the trajectories with low $E_k$ are considered in the ensemble averages. 
 
Considering an idealized puff,  whose dynamic behaviors are only governed by the deterministic components, maintains a steady state,  then we have $\sigma_d=0$ or $\mathcal{D}_d(E_k)/\mathcal{P}_d(E_k)=Rs$, where $Rs$ is the corresponding Reynolds number for such a state with a disturbance kinetic energy of $E_k$. The critical Reynolds number  is the minimum of $Rs$, 
\begin{equation}
	Re_c=\frac{ \mathcal{D}_d}{ \mathcal{P}_d}, \text{ where } \frac{d( \mathcal{D}_d/ \mathcal{P}_d)}{dE_k}=0 \text{ or } \frac{d \mathcal{D}_d}{d \mathcal{P}_d}=\frac{ \mathcal{D}_d}{ \mathcal{P}_d}.
	\label{eq:eq7}
\end{equation}
Based on the  $\mathcal{P}_d$ and $\mathcal{D}_d$ obtained at low Reynolds numbers, the critical parameters $(Re_c, E_{kc})=(1445, 0.355)$ can be predicted with Eq.~(\ref{eq:eq7})  as shown in Fig. 4(c), where $d\mathcal{D}_d/d\mathcal{P}_d$ is calculated with the quintic fitting  curves  [Fig. 4(a) and 4(b)], and  agree well with 1450, the $Re_c$ obtained from the -2 scaling law (Eq. 4).

\begin{figure}[b]
	\centering
	\includegraphics[width=0.77\linewidth]{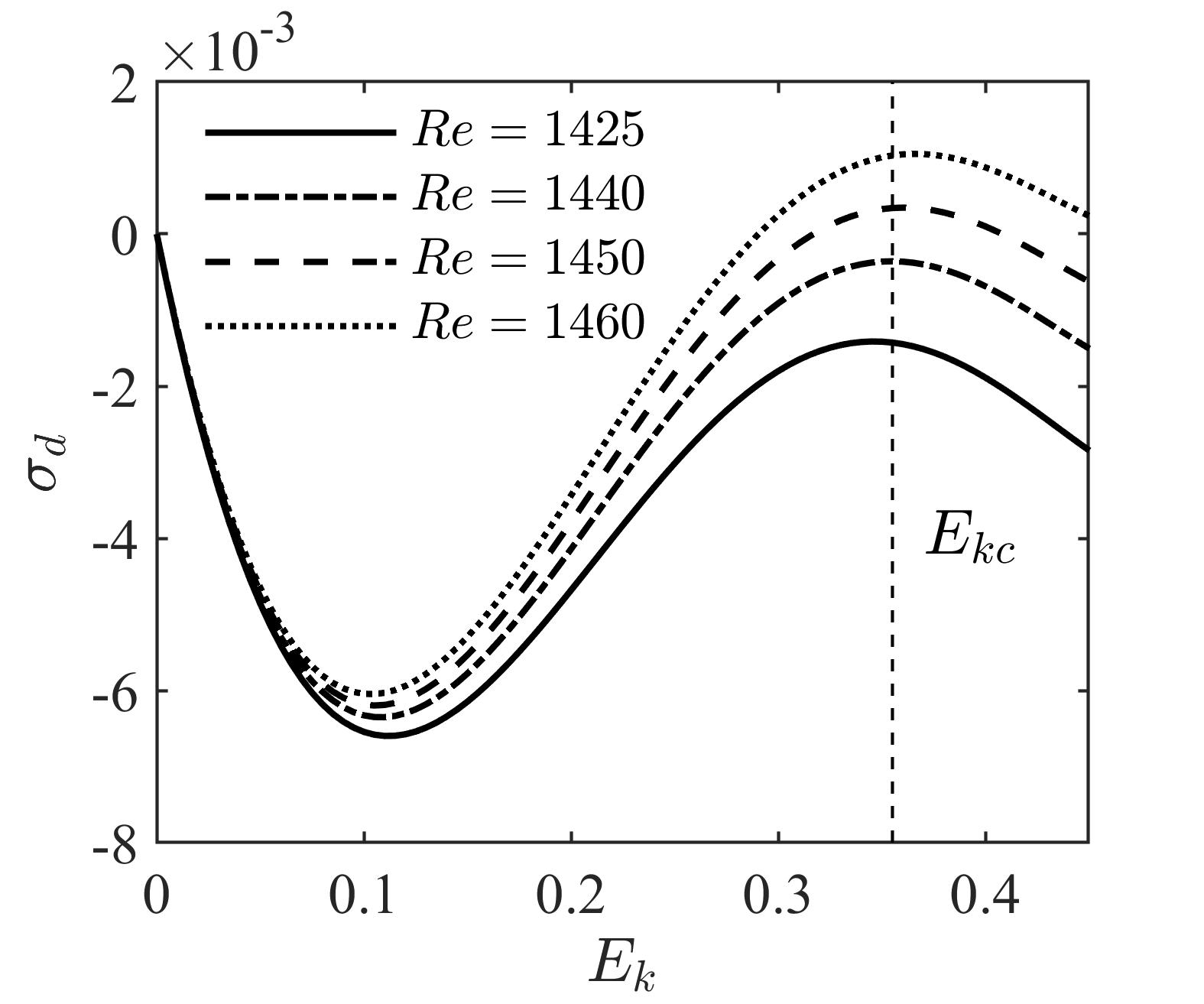}
	\caption{ $\sigma_d$ as a function of $E_k$ at different Reynolds numbers. }
	\label{fig:f5}
\end{figure}

When ${|Re_c-Re|}/{Re_c}\ll 1$, it is shown in Fig.~\ref{fig:f5} that  the deterministic growth rates reach their maxima around $E_k=E_{kc}$, i.e., $d \sigma_d/dE_k\simeq 0$, and $[d^2 \sigma_d/dE_k^2]_{E_k=E_{kc}}$is close to $ [d^2 \sigma_d/dE_k^2]_{(Re,E_k)=(Re_c,E_{kc})}=-A<0$. The growth rate can be expanded at $E_k=E_{kc}$ as
\begin{equation}
	\begin{split}
\sigma_d(E_k)&\simeq \sigma_d(E_{kc})  -\frac{A}{2}(E_k-E_{kc})^2+... \\
 &\simeq   \mathcal{P}_d(E_{kc})-\frac{\mathcal{D}_d(E_{kc})}{Re}-\frac{A}{2}(E_k-E_{kc})^2\\
  &\simeq   \mathcal{D}_d(E_{kc}) \frac{Re-Re_c}{ReRe_c} -\frac{A}{2}(E_k-E_{kc})^2\\
   &\simeq  \mathcal{D}_d(E_{kc}) \frac{Re-Re_c}{Re_c^2} -\frac{A}{2}(E_k-E_{kc})^2.
\end{split}
\end{equation}
Substituting the above equation into Eq. (6), we have
\begin{equation}
	\frac{d(E_k-E_{kc})}{dt}\simeq\mathcal{D}_d(E_{kc}) \frac{Re-Re_c}{Re_c^2} -\frac{A}{2}(E_k-E_{kc})^2+\sigma_s.
\end{equation}
This is the classical equation governing a stochastic or noisy saddle-node bifurcation \cite{Hathcock2021}. 

When $Re<Re_c$, Eq. (8) indicates $\sigma_d<0$, indicating that puffs will deterministically decay. It is known that in the ghost region of a classical saddle-node bifurcation, the nonlinear dynamic system  exhibits critical slowing down phenomenon \cite{Manneville2004,Strogatz2019}, yielding the square-root scaling law, which corresponds to Eq. (4), i.e.,  $\tau \propto (Re_c-Re)^{-1/2}$. In fact, we may rescale the time and the reduced kinetic energy with the mean lifetime $\tau$ as $t={\tau}\tilde{t}$, $E_k -E_{kc} =2\tilde{E_k}/(A{\tau} )$, and $\sigma_s=2\tilde{\sigma_s}/(A\tau^2)$, Eq.~(9) can be transformed into the unified form $\mathrm{d}\tilde{E}_k /\mathrm{d}\tilde{t}=-1-\tilde{E}_k^2+\tilde{\sigma_s}$ for the decay cases by setting $1/{\tau}^2=(\rm{Re}_c-\rm{Re})A \mathcal{D}_d(E_{kc})/(2\rm{Re}_c^2)$, i.e. ${\tau} \propto (\rm{Re}_c-\rm{Re})^{-1/2}$, the lifetime's square-root scaling law found in DNS and shown  in Fig. 3(a). This lifetime scaling law of transitional structures has been found for two-dimensional channel flows \cite{Zhang2022} and pipe flows \cite{guan2025}, suggesting a universal property for the subcritical transitions of wall-bounded shear flows.

When $Re$ is larger than $Re_c$, the deterministic growth rate $\sigma_d$ becomes positive around $E_{kc}$ as shown in Fig. 5. The bifurcation [Eq. (8)] intrinsically defines a stable node (upper) branch in the phase space, persisting the turbulent states and leading to longevity, and a unstable saddle (lower) branch, providing  a potential energy barrier. According to the nonlinear dynamic theory, the stochastic fluctuations, represented by the $\sigma_s$ term in Eq. (9), enable a barrier crossing process, which leads to memoryless decay or abruptly relaminarization  \cite{Hathcock2021, guan2025}, a key feature of metastable state \cite{Barkley2016}, and super-exponential growth of mean lifetime with $Re$. Therefore, $Re_c$ revealed in this letter is the critical Reynolds number for metastable turbulent puffs in duct flows.   When $Re$ is increased further, puff splitting is observed \cite{Khan2021} and the balance between puff decay and splitting will be achieved, suggesting a turbulence persistence pathway analogous to that in pipe flows \cite{Avila2011}.  

Localized turbulence is a key feature of the subcritical transitions in wall-bounded shear flows. Different from the axisymmetric  pipe flows, turbulent puffs in square duct exhibit distinctive secondary flows.  Ensemble-averaged fields reveal organized spatial patterns: (1) corner-aligned localized high-speed streaks flanked by streamwise vortex pairs, (2) a central low-speed streak bifurcating into four near-wall branches at the upstream front. Despite these structural modifications, the transient nature of puffs remains qualitatively unchanged from the pipe flow counterparts. Based on direct numerical simulations, the Reynolds-Orr equation, and the pattern preservation approximation, we establish that the puff dynamics obey the noisy saddle-node bifurcation scenario. Below the critical Reynolds number $Re_c$, deterministic decay governed by the critical slowing down dictates the observed square-root lifetime scaling, while supercritical regime ($Re>Re_c$) exhibits metastability, where the stochastic-fluctuation activated barrier crossing process produces the abrupt decay and the super-exponential lifetime dependence. These findings are expected to advance the theoretical framework for transitional shear flows by reconciling structural diversity with unified dynamical principles, offering predictive scaling laws for potential applications involving duct geometries.

\begin{acknowledgements}
	The support from the National Natural Science Foundation of China is acknowledged (Grants No. 91752203).
\end{acknowledgements}
\section*{Data Availability}
The data that support the findings of this study are available within the article.

\nocite{*}
\bibliography{duct}

\end{document}